\documentclass{amsproc}

\newtheorem{theorem}{Theorem}[section]

\theoremstyle{definition}

\theoremstyle{remark}

\numberwithin{equation}{section}

\newcommand{\be}{\begin{equation}}
\newcommand{\ee}{\end{equation}}
\newcommand{\e}{\epsilon}

\newcommand{\R}{\mathbb R}

\newcommand{\Probabilities}{\mathcal P }
\newcommand{\esssup}{\mathop{\rm ess \, sup}}
\newcommand{\PP}{\mathbb P}
\newcommand{\mathscr}{\mathcal }
\newcommand{\mmu}{{\mbox{\boldmath$\mu$}}}
\newcommand{\Vtild}{\widetilde V}

\begin{document}

\title[Rough potentials]{Recent results in semiclassical approximation with rough potentials}  

\author{T. Paul}
\address{CNRS and CMLS, \'Ecole polytechnique, 91128 Palaiseau cedex, FRANCE}  
\curraddr{} 
\email{paul@math.polytechnique.fr}



\maketitle


\small
Quantum Mechanics was invented for stability reasons. In fact it is striking to notice the difference of regularity that needs the potential of a Schr\"odinger operator
to insure unitary of the quantum flow 
(e.g. $V\in L^1_{\mbox{loc}},\ \ \lim_{\e\to 0}\sup_{x}\int_{\vert x-y\vert\leq\e}
\vert x-y\vert^{2-N}\vert V(y)\vert dy=0$) compared to the classical Cauchy-Lipshitz condition for vector fields.

On the other hand, tremendous progress have been done in the last 25 years concerning the theory of ODEs using PDE's methods: 
extension of the
Cauchy-Lipshitz condition to Sobolev ones (DiPerna-Lions 1989) and BV vector fields 
(Bouchut for the Hamiltonian case 2001 and Ambrosio for the general case 2004) have been proved to provide well-posedness of the classical flow almost everywhere, through
uniqueness result for the corresponding Liouville equation in the space $L^{\infty}_+([0,T];L^{1}(\R^{2n})\cap L^{\infty}(\R^{2n}))$.
Under these regularity conditions on the potential (in addition to some growing at infinity) the Schr\"odinger equation is well posed for all positive 
values of the Planck constant and it is therefore natural to ask what is happening  at the classical limit. 
As we will see, different answers will be given, according to the choices we make first on the topology of the convergence, 
and secondly on the asymptotic properties of the initial datum. 
The general idea of the results we are going to present here can be summarized as follows:
\vskip 0.3cm
For some $ V\notin C^{1,1}$ both the quantum  and the classical exist and
\begin{center}
\textit{ the diPerna-Lions-Ambrosio flow is the classical limit of the quantum flow}

for non concentrating initial data.

For concentrating initial data

\textit{the multivalued bicharacteristics are the classical limit of the quantum flow.}

\end{center}
\vskip 0.3cm
All the results presented here will use a quantum formalism on phase space, thanks to the notion of Wigner function. 
More precisely we will be concerned with the so-called Schr\"odinger and von Neumann equation
\[
i\e\partial_t\psi=(-\e^2\Delta+V)\psi\ \mbox{ and }\ \partial_t D=\frac 1 {i\e}[-\e^2\Delta+V,D]
\]
 with   $\psi^{t=0}\in L^2(\R^n)$ and $D^{t=0}\geq 0,
TrD^{t=0}=1$ (density matrix, e.g. $D^0=\vert\psi^0\rangle\langle\psi^0\vert$). 
And we will consider the Wigner function associated to $D^t$ (e.g. $=\vert\psi^t\rangle\langle\psi^t\vert$), defined by
\[
W^\e D(x,p):=\frac{1}{(2\pi)^n}\int_{\R^n}D^t(x+\frac{\e}{2}yx-\frac{\e}{2}y)e^{-ipy}dy
\]
where $D^t(x,y)$ is the integral kernel of $D^t$ (e.g. $=\overline{\psi^t(x)}\psi^t(y)$ in which case we write $W^\e\psi^t_\e$).

The well-known lack  of positivity of $W^\e$ suggests, in order to study evolution in spaces like $L^{\infty}_+$, to use the so-called Husimi function
of $D^t$, a mollification of $W^\e$ defined as $\widetilde{W^\e D}:=e^{\e\Delta_{\R^{2n}}} W^\e D$ which happens to be positive. 
But the only bound we have for $\widetilde{W^\e D}$ is
$\Vert\widetilde{W^\e D}\Vert_{L^\infty}\leq\e^{-n}\mbox{Tr}D$, unuseful  for the $L^\infty$ condition needed
for the existence of the classical solution. We formulate the

\vskip 0.3cm
\textbf{Conjecture}:\ For an $\e$-dependant family $D_\e$ of density matrices we have
\[ \mbox{Tr}D_\e=1 \Longrightarrow \sup\limits_{\e>0}\Vert\widetilde{W^\e D_\e}\Vert_{L^\infty}=+\infty.\]

In the general case of a potential whose gradient is  BV, the first idea will be to smeared out the initial conditions 
and consider a family of vectors $\psi^0_{\e,w},\ w$ belonging to a probability space $(W,{\mathcal F},\mathbb P)$. Under 
the general assumptions 

\scriptsize
\[
\begin{array}{ccc}
\mbox{\textbf{Assumptions on }} V &&\mbox{\textbf{Assumptions on initial datum}}\\
globally\  bounded,\ locally\ Lipschitz&&\psi^\e_{0,w}\in H^2(\R^n;\mathbb C)\\
\nabla U_b\in BV_{{loc}}(\R^n;\R^n)&&\sup_{\e>0}\int_W\int_{\R^n}|H_\e\psi^0_{\e,w}|^2\,dx\,d\PP(w)<\infty\\
\esssup_{x \in \R^{n}}\, \frac{|\nabla U_b(x)|}{1+|x|} <+\infty&&\int_W\vert\psi^0_{\e,w} ><\psi^0_{\e,w} \vert\,d\PP(w)\leq \e^n{\rm Id}\\
+ finite\  repulsive\ Coulomb\ singularities&&
\lim_{\e\downarrow 0}\widetilde{ W^\e\psi^0_{\e,w}}=i(w)\in\Probabilities{(\R^d)}\mbox{ for }\PP-a.e.\ w\in W.

\end{array}\]
\small
we have, for any bounded distance $d_{{\mathcal P}}$ inducing the weak topology in $\mathcal P({\mathbb R^{2n}})$, the
\begin{theorem}[\cite{AFFGP}]\label{1}
\begin{equation}\nonumber
\lim_{\e\to 0} \int_W\sup_{t\in [-T,T]}d_{{\mathscr
P}}\bigl(\widetilde{W^\e\psi^t_{\e,w}}),\mmu(t,i(w))\bigr) \,d\PP(w)=0,
\end{equation}
where $\mmu(t,\nu)$ is a (regular Lagrangian) flow on $\Probabilities{(\Probabilities{(\R^{2n}))}}$ ``solving" the Liouville equation.
\end{theorem}
 In the case of the von Neumann equation, a more direct result can be obtained.
 \scriptsize
 \[
\begin{array}{ccc}
\mbox{\textbf{Assumptions on }} V &&\mbox{\textbf{Assumptions on initial datum}}\\

globally\  bounded,\ locally\ Lipschitz&&sup_{\e\in (0,1)}\mbox{Tr}(H_{\e}^2D^o_\e) <+\infty\\
\nabla U_b\in BV_{{loc}}(\R^n;\R^n)&&D^o_\e\leq\e^n\mbox{Id}\\
\esssup_{x \in \R^{n}}\, \frac{|\nabla U_b(x)|}{1+|x|} <+\infty&&w-\lim_{\e \to 0}  W^\e 

D^0_\e=W^0_0\in \mathcal P{\R^{2n}}\\

\end{array}\]\small
 
\begin{theorem}[\cite{FLP}]\label{2}
Let $d_{{\mathcal P}}$ be any bounded distance inducing the weak topology in $\mathcal P({\mathbb R^{2n}})$. Then
\begin{equation}\nonumber
\lim_{\e \to 0} \sup_{[0,T]} d_{{\mathscr P}}(\widetilde{W^\e D^t_\e},W^0_t )=0,
\end{equation}
$W^0_t$ is the unique solution in $L_+^{\infty}([0,T];L^1(\R^{2n})\cap L^{\infty}(\R^{2n}))$ of the Liouville equation.
\end{theorem}

The next result concerns the semiclassical approximation in strong topology. Let us denote $\widetilde V:=e^{\e\Delta_{\R^n}}V$ and suppose:
\scriptsize
\[
\begin{array}{ccc}
\mbox{\textbf{(new) Assumptions on }} V &&\mbox{\textbf{(new) Assumptions on initial datum}}\\

\int{|\widehat{V}(S)| \,\, \frac{\vert S\vert^2}{1+\vert S\vert^2} \,\,dS} < \infty &&W_0^\e\in H^2(\mathbb R^n)\\
\int\limits_{|S|\in (a,b)}{|\widehat{V}(S)| \, |S|^m dS} 
\leq C\left(b^{m-1-\theta}-a^{m-1-\theta}\right) &&\mbox{if }\partial_t \rho+ k \partial_x \rho- \partial_x \Vtild \cdot \partial_x \rho =0,\ \rho_{t=0}:=W^\e_0\\
m=0,1,2,\ 0<\theta<1&&\exists T>0$, $\delta \in (0,\frac{\theta}{2+\theta})\mbox{ such that}\\
&&||\rho(t)||_{H^2} =O(\varepsilon^{-\delta}||W_0^\varepsilon||_{L^2})$ for $t\in[0,T]
\end{array}\]\small

\begin{theorem}[\cite{apbad}]\label{3}
Let $\rho_1^\e$ be the solution of
\begin{equation}
\nonumber
 \partial_t \rho_1^\e+ k \partial_x\rho_1^\e- \partial_x \Vtild \cdot \partial_x \rho_1^\e =0.
\end{equation}
$W^\e_t:=W^\e D^t_\e$   satisfies, uniformly on $[0,T]$,
\begin{equation}
\nonumber
|| W^\e_t-\rho_1^\varepsilon(t)||_{L^2} =O(\varepsilon^{\kappa}||W_0^\varepsilon||_{L^2}),\ 
\kappa=min\{ \,\,\frac{1+\theta}{2}-1, \,\, \frac{\theta}{2+\theta}-\delta \,\,\}.
\end{equation}

\end{theorem}
 Let us finally give a 1D example where the lack of unicity will be crucial. 
 
 Let $V$ be a confining potential such that
$V=-\vert x\vert^{1+\theta}$ near $0$.
Near $(0,0)$ we obtain two solutions of the Hamiltonian flow:
\begin{equation}\nonumber
(X^\pm(t), P^\pm(t)) = ( \pm c_0 t^\nu ,\pm{c_0\nu} t^{\nu-1}  ),\ \nu=\frac{2}{1-\theta} \mbox{ and } 
c_0=\left({  \frac{(1-\theta)^2}{2} }\right)^{1-\theta},
\end{equation}
 plus a continuum family of solutions obtained by not moving up to any value $t_0$ of the time and then starting to move according to $(X^\pm(t-t_0), P^\pm(t-t_0))$. 
 
 The question now is to know which one, out of this continuous family, is going to be selected by the classical limit. The answer is given by the following result.
 
 \begin{theorem}[\cite{apbad}]\label{4}
 Let $W^\varepsilon D^0_\e(x,k)=\lambda^{\frac{7+3\theta}{30}}w( {\lambda^{\frac{1+\theta}{6}}}x,
 {\lambda^{\frac{1-\theta}{15}}}k),\\ \lambda=\log{\frac1\e},\ 
 supp\, w \subseteq \{ |x|^2+|k|^ 2<1 \}$. 
 
 Then
$\exists T>0$ s.t.  $\forall t\in[0,T]$,  $W^\e D_\e^t$ 
converges in weak-$*$ sense  to
\be\nonumber
W^0_t=c_+\delta_{(X^+(t),P^+(t))}+c_-\delta_{(X^-(t),P^-(t))}, \mbox{ with }
c_\pm = \int\limits_{\pm x >0}{w(x,k)dxdk}.
\ee
\end{theorem}
 
 What these results show is the fact that, at the contrary to the case where the underlying classical dynamics is well-posed, the classical
 limit of the quantum evolution with non regular (i.e. not providing uniqueness of the classical flow) potentials is not unique, and depends on
  the family  of initial conditions itself, and not anymore only on their limit.
  
  For non concentrating data the classical limit, in the general case of a potential whose gradient is BV, is driven (in the two senses expressed by Theorems \ref{1} and \ref{2})
  by the DiPerna-Lions-Bouchut-Ambrosio flow.
  
  Slowly concentrating data (Theorem \ref{4}) provide situations where the classical limit is ubiquitous, 
  and follows several of the non unique bicharateritics, a typical quantum feature surviving, in this situation, the classical limit. 
  It is important to remark that
  the speed of concentration governs the selections of the remaining trajectories. The case of fast concentration, in particular the pure states situations, is still open.

\end{document}